\documentclass[aps,twocolumn]{revtex4}
\usepackage{graphics}

\begin{document}
\bibliographystyle{prsty}
\def\Lya{Ly$\alpha\ $}

\title{Cosmology with the Ly-$\alpha$ forest}
\author{Martin White}
\affiliation{Dept Physics and Dept Astronomy, University of California,
  Berkeley, CA 94720}
\date{\today}

\begin{abstract}
The \Lya forest has emerged as one of the few systems capable
of probing small-scale structure at high-$z$ with high precision.
In this talk I highlight two areas in which the \Lya forest is
shedding light on fundamental questions in cosmology, one speculative
and one which should be possible in the near future.
\end{abstract}

\pacs{98.70.Vc}
\maketitle

\section{Introduction}

As we heard in the talks by Uros Seljak and Lam Hui in this meeting,
the physics of the \Lya forest, as measured in high-$z$ Quasi-Stellar
Object (QSO) spectra, is reasonably simple.  At these redshifts the gas making
up the intergalactic medium is in photoionization equilibrium, which results
in a tight density--temperature relation for the absorbing material with the
neutral hydrogen density proportional to a power of the baryon density.
Since pressure forces are sub-dominant, the neutral hydrogen density closely
traces the total matter density on the scales relevant to the forest
($0.1-10\,h^{-1}$Mpc).
The structure in QSO absorption thus traces, in a calculable way, slight
fluctuations in the matter density of the universe back along the line of
sight to the QSO, with most of the \Lya forest arising from over-densities
of a few times the mean density.

In this contribution I want to focus on two areas where the \Lya forest
can be used to constrain cosmology and astrophysics.  The first, dealing
with measuring the matter power spectrum at high redshift, is quite
speculative, but might be very exciting.  The second has a higher chance
of success and may tell us something about the nature of the ionizing
sources, the end of the dark ages and the formation of the first objects
in the universe.

\section{Finding baryons at $z>1$?}

The presence of a series of almost regularly spaced peaks in the cosmic
microwave background temperature angular power spectrum has now been
compellingly demonstrated.  The same physics which gives rise to these
peaks, acoustic oscillations in the baryon-photon plasma prior to
recombination \cite{PeeYu}, also predicts a (much smaller) series of peaks
in the matter power spectrum.
A measurement of these peaks would provide a fundamental test of the paradigm,
and an independent measure of the $z$-dependent Hubble constant and
distance-redshift relation
\cite{EisHuSilSza,EisHuTeg,MeiWhiPea,Eisenstein,BlaGla,Lin}.
Because non-linear growth `washes out' the peaks \cite{MeiWhiPea} they are
best searched for at high redshift.

\begin{figure}
\begin{center}
\resizebox{3in}{!}{\includegraphics{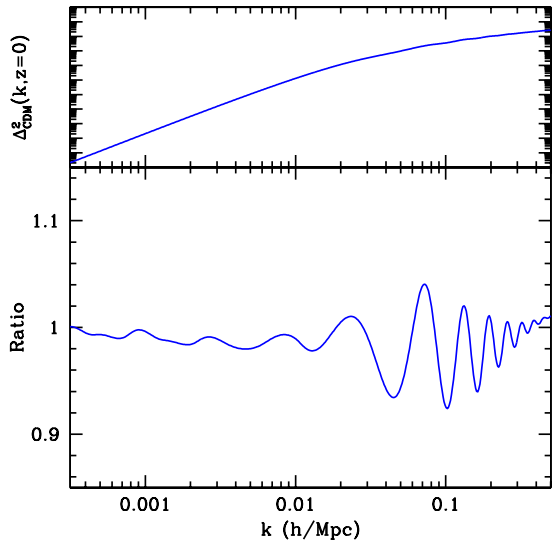}}
\end{center}
\caption{The contribution to the dark matter variance per $\log k$,
$\Delta^2(k)$, in linear theory.
The upper panel shows $\Delta^2(k)$ on a logarithmic scale while the lower
panel shows the ratio of the numerically computed $\Delta^2(k)$
\protect\cite{WhiSco} to the fitting function of \protect\cite{EisHu} with
the oscillations suppressed.}
\label{fig:d2}
\end{figure}

The baryon oscillations themselves lie at $k\simeq 0.03-0.3\,h/$Mpc
(see Fig.~\ref{fig:d2}),
with an amplitude of 5\% or less for currently popular cosmological
models \cite{EisHuSilSza,MeiWhiPea}.
The higher modes are roughly out of phase with the photon peaks
(also in $k$-space) and appear half as often as the peaks in the
CMB spectrum (see e.g.~Fig.~1 of \cite{MeiWhiPea}).
The peaks are suppressed by Silk damping beyond $k\sim 1\,h/$Mpc.

Naively the best way to measure the small baryonic oscillations in the
matter power spectrum would be an ultra-large galaxy redshift survey at
high-$z$.  Such a survey would require next-generation instruments and a
large dedicated team.  Thus even with adequate funding and heroic efforts
it is several years away.  It is therefore interesting to explore whether
one could measure the baryonic oscillations another way.

Within a few years the SDSS will return a catalogue of $\sim 10^5$ QSOs
over $\sim 10^4$ square degrees of sky.
These quasars will be distributed in both redshift and angular separation,
but simple counting suggests that one should find many hundreds to several
thousands with separations of tens of Mpc above $z\sim 2$.
It is therefore interesting to ask whether this sample could be used to
probe the baryon oscillations.

Using simply the auto-power spectrum of the forest will not work, because
the line-of-sight spectrum at $k$ is an integral of the 3D spectrum over
wavenumbers greater than $k$.  This integration suppresses the small
oscillations which are our primary signal.  However one can consider the
cross-spectra, where one line-of-sight is cross-correlated with a
neighboring line-of-sight.  As we shall show below this statistic is
sensitive to the oscillations.  In addition uncorrelated errors from
e.g.~continuum fitting, should be averaged down in this procedure relative
to their value in the auto-spectrum.  Since we are working on large scales,
where the noise is essentially absent, one could in principle measure
the flux cross-spectrum to the cosmic variance limit
\begin{equation}
  {\delta \pi_\tau\over\pi_\tau} \simeq
    \sqrt{2 k_f\over N \Delta k}
\end{equation}
where $N$ is the number of pairs, $k_f$ is the fundamental mode along the
line-of-sight and $\Delta k$ is the width of the bin in $k$-space in which
we are measuring the power.  We can alternatively write $k_f=2\pi/L$,
where $L$ is the length of the `useful' part of the spectrum.
Unless we attempt to fit to the higher lines, the useful length of the
spectrum is the section between \Lya and Ly$\beta$ which covers hundreds
of Mpc in the rest frame of the absorbers.
Thus with a few hundred pairs it is possible to measure the cross-spectrum
with error bars on the order of per cent.

One major advantage here is that, like the CMB itself, the redshift and
scale dependence of the peaks is accurately calculable given a cosmological
model.  This allows us to perform a likelihood estimation of the relevant
cosmological factors, including pairs of different distances and redshifts,
rather than attempt to perform an inversion on the irregularly sampled data.
Though the latter can be done, it is considerably more difficult than the
`forward' approach.

By applying the idea of the peak-background split from large-scale
structure is it possible to show that, for sufficiently large scales and
assuming $\tau=A(\rho/\bar{\rho})^\alpha$, the flux cross-spectrum,
$\pi_\tau$, along two lines of sight separated by distance $x_\perp$ is
proportional to the dark matter cross-power spectrum $\pi$
(see also \cite{McDonald,Viel}).
On these very large scales we can neglect the effects of thermal
broadening or peculiar velocities.
The DM cross-spectrum is simply related to the 3D power spectrum by
\begin{equation}
  \pi^2(k,x_\perp) = \int_k^\infty {dq\over q} \left({k\over q}\right)
     J_0(\sqrt{q^2-k^2})\Delta^2(q)
\end{equation}
or \cite{Viel}
\begin{equation}
  \Delta^2(q) = \int_0^\infty {dx\over x} \left({qx}\right)^2
     J_0(x\sqrt{q^2-k^2})\pi^2(q)
\end{equation}
where the inverse relation holds $\forall\ k<q$.
Note that for smooth spectra, $\pi^2(k)$ gets contributions from
$k<q<x_\perp^{-1}$.  As $x_\perp\to 0$, $J_0\to 1$ and we find the
auto-spectrum is the projection of the 3D spectrum as anticipated.

Converting from $h/$Mpc to s/km using
$H(z)=H_0\sqrt{\Omega_M(1+z)^3+\Omega_\Lambda}
  \sim 100\,{\rm km}/{\rm s}/{\rm Mpc}$
we find that the relevant spectral range is roughly $3\times 10^{-4}$s/km to
$3\times 10^{-3}$s/km.  This is quite large scale, roughly comparable to the
fundamental mode if we measure only from 100nm to 120nm.

\begin{figure}
\begin{center}
\resizebox{3in}{!}{\includegraphics{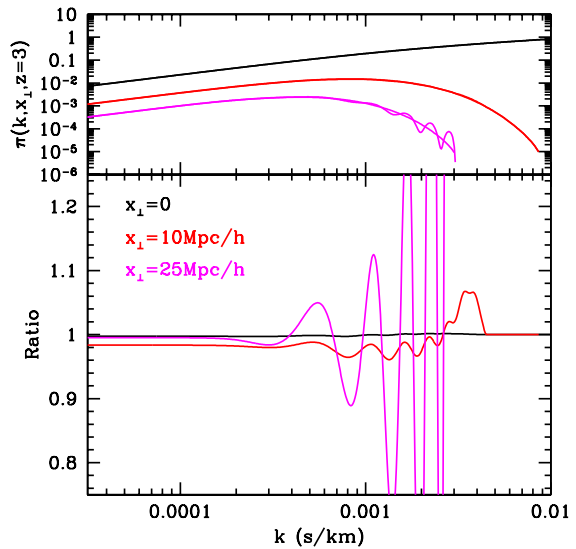}}
\end{center}
\caption{The cross-power spectra of the DM, in linear theory, at $z=3$.
The upper panel shows $\pi(k,x_\perp)$, including and excluding the baryonic
oscillations, for different values of $x_\perp$.
The lower panel shows the ratio to the smooth fitting function of
\protect\cite{EisHu}.}
\label{fig:pi}
\end{figure}

On the positive side Fig.~\ref{fig:pi} shows that the signal is present in
the cross spectra at separations of 10Mpc or so.
The relative amplitude of the oscillations (which is the figure of merit for
cosmic variance dominated measurements) can be quite large, 10\% or more,
compared to the accuracy with which $\pi$ can be measured.
Because we can accurately predict the positions and amplitudes of the peaks
we know both the $x_\perp$ and $z$ dependence of the signal.

On the negative side there is only a small fractional change in $\pi$
when $\pi$ is big (i.e.~$x_\perp$ small) and the larger fractional change
in $\pi$ when $\pi$ is small.  This suggests we would be working primarily
at very low signal levels, where systematics become increasingly important.
Also, if we work with the spectra only between \Lya and Ly$\beta$ we do
not have very much `spare' resolution in $k$ to see the oscillations.  We
would ideally probe $k\ll 0.003$s/km which requires very long spectra.

In summary, the signal of baryon oscillations is lurking in the cross-spectra
of QSO pairs, which allow us to probe large volumes of space at high redshift
with a modest investment in telescope time.  A naive estimate of the
cosmic variance limitations of a QSO survey like the SDSS suggests that
the signal could be seen in a likelihood analysis with high confidence.
However the overall signal levels are very low and there are numerous sources
of systematic error which may make the signal forever unmeasurable.

\section{The nature of the ionizing sources}

Currently the sources that make up the UV background are unknown.  However
the mean optical depth $\bar{\tau}$ in the \Lya forest offers a way to
constrain the source emissivity and fluctuations in the UVbg can be used to
constrain the nature of the sources.
I have been involved in a long-running collaboration with Avery Meiksin to
investigate these issues \cite{PaperI,PaperII}, and present some of our
basic results here.

First we consider the constraints on the source emissivity $\epsilon$.
Given a cosmological model, the mean flux or optical depth $\bar{\tau}$
fixes the ionization rate $\Gamma$.  If we approximate $\exp[-\tau]$ as
$\exp[-\ell/r_0]$ where $r_0$ is the {\it attenuation length\/} then
$\Gamma\sim \epsilon r_{\rm eff}$.  Here $r_{\rm eff}$ is an effective
attenuation length which interpolates between the attenuation limited
($r_0\ll\ell_H$) and cosmological expansion+age limited ($r_0\gg\ell_H$)
regimes and $\ell_H$ is the Hubble distance.

As a function of $\epsilon$ the transition between the attenuation
limited and cosmologically limited regimes is very abrupt, and at
fixed $\epsilon$, $r_0$ falls rapidly with increasing redshift.
This in turn means that $\bar{\tau}$ rises rapidly, almost exponentially,
with redshift.

\begin{figure}
\begin{center}
\resizebox{3in}{!}{\includegraphics{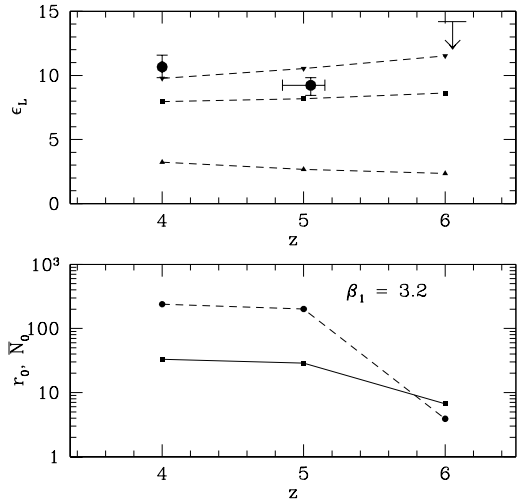}}
\end{center}
\caption{(Upper) The emissivity, at the Lyman edge, required to fit the mean
flux observed in QSO spectra as a function of wavelength (points) compared
to extrapolated QSO emissivities for 3 different assumptions about the
evolution of the QSO luminosity function (lines).  Under these assumptions
QSOs would be enough to keep the universe ionized at $z\le 6$.
(Lower) The attenuation length, $r_0$ in proper Mpc/$h$, as a function of $z$.
Note that $r_0$ drops quasi-exponentially to $z\sim 6$.}
\label{fig:fig3}
\end{figure}

Fig.~\ref{fig:fig3} shows the emissivity of the sources, as a function
of $z$, required to fit the observed $\bar{\tau}$.  This is compared to
extrapolations of the QSO luminosity function.
From this we can see that QSOs may contribute a substantial fraction of
the UV background even as high as $z\sim 6$.
If QSOs do contribute non-negligibly to the UV background then we would
expect large spatial fluctuations in $\Gamma$.  Conversely, if galaxies
or star clusters dominate the UV background we expect to see only the
imprint of large-scale structure.
Unfortunately the effects of UV background correlations are most
conspicuous at $z>5$, where the precise level of absorption by the \Lya
forest is most difficult to measure because of extremely low flux values.

\begin{figure}
\begin{center}
\resizebox{3in}{!}{\includegraphics{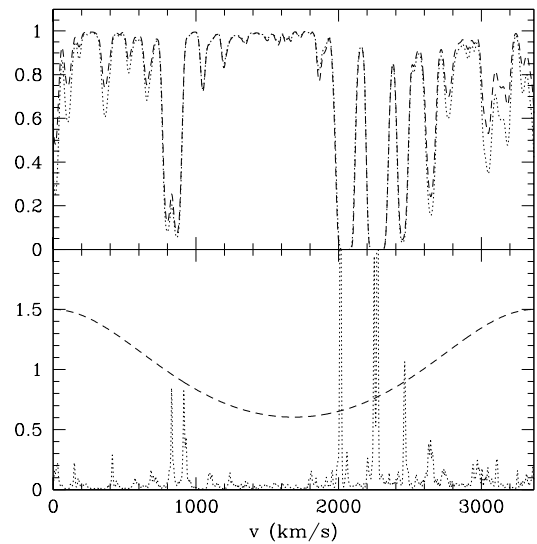}}
\end{center}
\caption{(Upper) The normalized flux, as a function of velocity, from a PM
simulation of structure formation.  The spectra were computed as described
in \protect\cite{PaperI,PaperII} assuming either a uniform ionizing
background (dashed) or that the background was dominated by a single source
at the center of the box (dotted).
(Lower) The dark matter density (dotted) along the line-of-sight or the
inverse ionization rate, $\Gamma^{-1}$, in units of the average (dashed).}
\label{fig:onesource}
\end{figure}

Fig.~\ref{fig:onesource} shows how a simulated \Lya spectrum is changed
if the UV background is (unrealistically) dominated by a single source
as compared to the usually assumed smooth behavior.
To assess a more realistic situation it is necessary to resort to Monte
Carlo simulations in which sources are placed within the simulation volume,
the UV background computed and then spectra created from the density and
velocity fields within the box.  Our investigations suggest that the 1-point
distribution is relatively unaffected by fluctuations in the UVbg, however
the 2-point statistics (the flux correlation function and power spectrum)
show noticeable differences at high redshift \cite{PaperII}.
In addition fluctuations in the UVbg will increase the estimated \Lya
optical depth over the value that would be estimated assuming a
homogeneous background \cite{PaperI}.
This increases the demand placed on the ionization rate required to
reproduce $\bar{\tau}$ and needs to be included when comparing with
estimates of source emissivity.

\section{Conclusions}

The \Lya forest is emerging as a powerful probe of structure formation
at high redshift.  Because the physics is relatively well understood we
can compare increasingly realistic calculations with ever better data
to constrain our cosmological models.  Here we have investigated two ways
in which the \Lya forest constrains cosmology.  The first, speculative,
proposal was to search for baryonic oscillations in the matter power spectrum
through their imprint in the \Lya forest cross-spectra.  The second proposal
was to constrain the emissivity of the sources of the UV background using
the measured mean optical depth of QSO spectra.

In the first case we found that while in principle the signal-to-noise and
statistics of the signal were favorable, in practice the very low signal
levels made this measurement prone to numerous systematic effects.
In the second case we showed that reasonable extrapolations of the QSO
luminosity function to high-$z$ would allow QSOs to contribute a
non-negligible fraction of the UV background even as early as $z=6$.
If this turns out to be the case, the UV background should have relatively
large spatial fluctuations which can affect the 2-point statistics of the
forest at high $z$.

\noindent {\it Acknowledgements:\/}
I thank Simon White for allowing me to present our work on baryon
oscillations in the \Lya forest, and Avery Meiksin for several years
of productive collaborations, including many enlightening conversations
on the physics of the IGM.  I would also like to thank the organizers
for such an interesting and successful conference.

\end{document}